\documentstyle[12pt,fleqn]{article}
\newcommand{\half}{\mbox{$\textstyle \frac{1}{2}$}}
\newcommand{\kett}[1]{\left | \, #1 \right \rangle}
\newcommand{\braa}[1]{\left \langle #1 \, \right |}

\newcommand{\beq}{\begin{equation}}
\newcommand{\eeq}{\end{equation}}
\newcommand{\beqa}{\begin{eqnarray}}
\newcommand{\eeqa}{\end{eqnarray}}

\def\ket{\rangle}
\def\bra{\langle}

\def\s{\sigma}

\def\sz{\sigma_z}

\def\sza{\sigma_z^a}
\def\s+a{\sigma_+^a}
\def\s-a{\sigma_-^a}

\def\szb{\sigma_z^b}
\def\s+b{\sigma_+^b}
\def\s-b{\sigma_-^b}

\def\w{\omega}
\def\wk{\omega_k}
\def\wo{\omega_0}
\def\woa{\omega_0^a}
\def\wob{\omega_0^b}

\def\ewt{e^{i \wk t}}

\def\ewtp{e^{i \wk t'}}
\def\emwtp{e^{-i \wk t'}}

\def\bk{{\bf k}}  

\def\bR{{\bf R}}
\def\br{{\bf r}}

\def\sumk{\sum_{\bk}}
\def\gk{g_{\bk}}
\def\gka{g_{\bk}^a}
\def\gkas{g_{\bk}^{a*}}
\def\gkb{g_{\bk}^b}
\def\gkbs{g_{\bk}^{b*}}

\def\Bk{b_\bk}
\def\Bks{b_\bk^\dagger}

\title{Quantum Computers and Dissipation}
\author{{\large G. Massimo Palma\protect\thanks {permanent address:  Istituto
di Fisica, via Archirafi 36, I-90123 Palermo, Italy},  Kalle-Antti
Suominen\protect\thanks{Theoretical Physics Division, Dept. of Physics,
University of Helsinki, P.O.BOX 9 FIN-00014,  Helsingin yliopisto,
Suomi-Finland}\ and Artur K. Ekert} \\ {\protect\small\em Clarendon Laboratory,
University of Oxford}\\ {\protect\small\em Parks Road, Oxford OX1 3PU, U.K.}}
\date{\today}

\begin{document}
\maketitle

\begin{abstract} We analyse dissipation in quantum computation and its
destructive impact  on efficiency of quantum algorithms. Using a general model
of decoherence we study the time evolution of a quantum register of arbitrary
lenght coupled with an environment of arbitrary coherence lenght. We discuss
relations between decoherence  and computational complexity and show that the
quantum factorisation algorithm  must be modified in order to be regarded as
efficient and realistic. 
\end{abstract}

\section{Introduction}

Quantum computers can accept input states which represent a coherent 
superposition of many different possible inputs and subsequently evolve them 
into a corresponding superposition of outputs. Computation, i.e. a sequence of 
unitary transformations, affects simultaneously each element of the 
superposition generating a massive parallel data processing albeit within one
piece  of quantum hardware.  As the result quantum computers can efficiently
solve some  problems which are believed to be intractable on any classical 
computer (Deutsch 1985, Deutsch and Jozsa 1992, Bernstein and Vazirani 1993, 
Simon 1994, Shor 1994). The most striking  example is the factoring problem: 
to factor a number $N$ of $L$ digits on any classical computer requires an 
execution time that grows exponentially with $L$ (approximately $\exp L^{1/3}$ 
for the best known algorithms such as the Number Field Sieve  (Lenstra {\em et
al.} 1990)).  In contrast Shor (1994) has shown that quantum computers require
an execution  time that grows only as a polynomial function of $L$ ($\approx
L^2$).   A noteworthy consequence for cryptology is the possibility of
breaking   public key cryptosystems  such as RSA (Rivest {\em et al.} 1979). 
In this context a practical implementation of quantum computation is a most 
important issue.

This paper shows that Shor's algorithm, in order to be efficient from the
computational complexity perspective, must be modified  (i.e. supplemented with
an exponentially efficient quantum error correction). This is because the error
rate, which is due to the interaction with the environment (a thermal
reservoir), grows as an exponential function of $L$ regardless the type and the
strength of the  computer - environment coupling. Consequently, in order to
obtain factors of $N$ with a prescribed probability of success one has to
repeat the computation $k$ times, where
$k$ grows exponentially with $L$. Let us stress that this observation is not
related to any technological limitations but is a consequence of the
computational complexity requirements imposed on any realistic model of
computation. Any technological progress in suppressing dissipation will, of
course, help to factor bigger and bigger numbers but will not change an
exponential increase of the error rate.

Our analysis is quite general and covers  cases where qubits interact with a
single reservoir and the distance between the qubits is comparable with the
correlation length of the reservoir.  In the limit of a short correlation
length (i.e. coupling to the independent reservoirs) our results coincide with
those by Unruh (1995).  Also we provide a  plausibility argument that any model
of dissipation will lead to errors that grow exponentially with $L$.

The structure of the paper is as follows: Section 2 links the computational
complexity of randomized algorithms with dissipation, Section 3  outlines basic
features of the dissipation model we adopted for this paper (the model is known
under the label ``decoherence"), Section 4 explains details of decoherence of a
single qubit using both semi-classical and fully quantised approaches, Section
5 extends the fully quantum analysis to the two qubit case  first and then to
the general case of an $L$-qubit register.  The results are summarised and
discussed in the last section.

\section{Errors in Randomized Algorithms}

An algorithm is said to be efficient if its execution time increases no  faster
than a polynomial function of the size of the input. For example, any 
efficient factoring algorithm will factor a number $N$ of size 
$L=\log N$ (the base of the logarithm is $10$ when the size is measured in
decimal digits and $2$ when it is measured in binary digits) in time
proportional to a polynomial function of $L$.  Algorithms do not have to be
deterministic;  random decisions can be taken at certain junctures in the
execution of the algorithm. The so-called {\em randomized algorithms} may
sometimes produce an incorrect solution, however; they have the property that
the probability of error can be made arbitrarily small.

Consider, for example, an efficient factoring algorithm which runs successfully 
only with probability $1-\epsilon$ ($0<\epsilon<1$). Clearly we know when the
algorithm is successful because we can check it efficiently by a trial
division.  Assuming that $\epsilon$ is {\em independent} of the input the
success probability of  the algorithm can be amplified (boosted) arbitrarily
close to $1$ by repeating the computation several times. After $k$ runs the
probability of having at least one success is 
\begin{equation} P(k,\epsilon) = 1-\epsilon^k .
\end{equation} If we fix  $P(k, \epsilon)$ (i.e. when we agree upon the desired
probability  of successful factorisation) then the number of runs $k$ depends
on the error  rate $\epsilon$. In order to retain efficiency of the algorithm
we may let 
$\epsilon$ increase with $L$ but not faster than a polynomial. Indeed,  Shor's 
quantum factoring algorithm is of this type; in a single run it provides
factors of $N$  with probability $1-\epsilon (L)$ and the probability of error 
$\epsilon (L)$ grows only as a polynomial function of $L$. However, errors in
Shor's algorithm are of purely ``mathematical" origin and correspond to the
random decisions which rely on the prime number theorem (i.e. on the
distribution of prime numbers). In a realistic scenario the probability of
error $\epsilon (L)$ has also contributions which are due to the computer -
environment coupling.   From the computational complexity point of view it is
irrelevant how weak this coupling is for a fixed
$L$, it is its dependence on $L$ which matters. If the interaction with the
environment causes an exponential increase of the error rate i.e. $\epsilon
(L)= 1-A\exp (-\alpha L)$, where $A$ and $\alpha$ are positive constants, then 
$k(L)\approx \exp(\alpha L)$ and the randomized algorithm cannot be regarded as
efficient any more!

Indeed, our results show that the computer-environment interaction leads to this
unwelcome exponential increase of the error rate.  Shor's algorithm, as it
stands at the moment, is not robust to the environmental noise and requires
modifications. This statement, however, should not be taken as a fundamental
argument against possibilities of experimental implementation of quantum
factorisation. It simply shows that further efforts, both theoretical (e.g.
designing an  exponentially efficient quantum error correction) and 
experimental (e.g. customizing the environmental noise), are necessary to make
quantum computation practical.

\section{Decoherence} 

Let us consider a quantum  register composed of $L$ qubits; each qubit being  a
two-state quantum  system described by a two-dimensional Hilbert space
${\cal H}_2$ with computational basis  states labeled as 
$\kett{0}$ and $\kett{1}$.  The $2^L$ dimensional Hilbert space of the whole
register is an $L$ term  tensor product of
${\cal H}_2$ pertaining to each qubit. Any quantum  state of the register can
be described  by a density operator of the form 

\begin{equation}
\rho (t) = \sum_{i,j=0}^{2^L-1} \rho_{ij}(t) \kett{i}\braa{j},
\end{equation} where the computational basis $\kett{i}$ is defined as a tensor
product of  the qubit basis states:

\begin{equation}
\kett{i} = \kett{i_{L-1}}\otimes\kett{i_{L-2}}\otimes\ldots 
\otimes \kett{i_{0}}.
\end{equation}

The r.h.s. is the binary decomposition of number $i=\sum_{l=0}^{L-1} 2^l i_l$,
e.g. for a two qubit register $\kett{2} = \kett{1}\otimes\kett{0}$.

Quantum computation takes its power from quantum interference and
entanglement.  The degree of the interference and entanglement in an $L$ qubit
register is  quantified by the coherences i.e. the off-diagonal elements of the
density  operator  in the computational basis $\rho_{ij}$ ($i\ne j$). When a
quantum computer is in contact with an environment, here modeled as a thermal
reservoir, the resulting dissipation destroys the coherences and changes the
populations (the diagonal elements).  When a reservoir is in a thermal state
the coherences  change on a much shorter time scale than populations and it is
decoherence which is responsible for destroying quantum parallelism. However,
it should be mentioned that in some cases populations can be more fragile than
coherences. The time scales depend on the type of the system-environment
coupling and on the quantum state of the reservoir (which does not have to be
thermal), for example, in some optically driven systems coherences change more
slowly than populations (Kimble \& Mabuchi 1995).

The notion of decoherence can be intuitively understood as follows. Consider  a
computer and a reservoir, initially both in pure states:

\begin{equation}
\kett{\Psi(0)} = \overbrace{\left(\sum_i
c_i(0)\kett{i}\right)}^{\mbox{computer}}
\otimes \overbrace{\kett{\alpha_0}}^{\mbox{reservoir}}.
\end{equation}

A unitary evolution of the composed system results in an entangled 
computer-reservoir state which can be written as

\begin{equation}
\kett{\Psi(t)} = \sum_i c_i(t)\kett{i}\otimes \kett{\alpha_i(t)},
\end{equation}

where, in general, $\braa{\alpha_i}\alpha_j\rangle \ne 0$ for $i\ne j$. The
elements of the register density matrix evolve as

\begin{equation}
\rho_{ij}(0)=c_i(0)c^*_j(0) \longrightarrow \rho_{ij}(t) =c_i(t)c^*_j(t)
\braa{\alpha_i(t)}\alpha_j(t)\rangle .
\end{equation}

In a popular model of decoherence (Zurek 1991), where the environment
effectively acts as a measuring apparatus, the evolution affects the reservoir
states
$\{\kett{\alpha_i}\}$ which become more and more orthogonal to each other
whilst the  coefficients $\{c_i\}$ remain unchanged. Consequently the
off-diagonal elements
$\rho_{ij}$ disappear due to the
$\braa{\alpha_i(t)}\alpha_j(t)\rangle$ factors. We will analyse this  type of 
decoherence in detail in the following sections.
 
The characteristic time for the off-diagonal elements to disappear is known as
the decoherence time. Its value depends on the type of qubits and  their
interaction with the environment and can vary from $10^4$ s for nuclear spins
in a paramagnetic atom to
$10^{-12}$ s for electron-hole excitations in a bulk of a semiconductor (for a
table of decoherence times for various physicsl systems see for  example
DiVincenzo 1995).

\section{Single Qubit Dephasing Mechanism}

In this section we want to describe in detail the basic features of the
decoherence model we adopted for this paper.  For this purpose it suffices to
analyse an interaction between a single qubit and the reservoir.  The case
involving two and more qubits will  be analysed in the subsequent sections.

The density operator  for a single qubit has the form
\begin{equation}
  \rho(t)=\sum_{i,j=0}^1 \rho_{ij}(t) \kett{i}\braa{j},
\end{equation} and we are interested in the time evolution of the coherence
$\rho_{01}(t)=\rho_{10}^{\star}(t)$.

\subsection{Semiclassical Picture} 

In a semiclassical analysis of the single qubit decoherence it is assumed  that
the reservoir is a fluctuating classical  magnetic field.  This approach
excludes many interesting quantum features, most notably the qubit-environment
entanglement;   however, we have decided to include it here because it can
simplify  {\em some}  calculations. 

We describe our physical qubit as a fictitious spin 1/2 system following  the
standard correspondence between the density operator $\rho$ and the spin 
vector $\vec s$, also referred to as the Bloch vector (Allen \& Eberly 1975, 
Cohen-Tannoudji {\em et al} 1977)

\begin{equation}
   \rho (t) = \frac{1}{2} (1 + \vec s(t)\cdot \vec\sigma ) = \frac{1}{2} 
   (1 +\sum_{a=x,y,z} s_a(t)\sigma_a),
\end{equation}

where $\sigma_a$, $a=x,y,z$ are the Pauli spin matrices. Components $s_x$ and
$s_y$ correspond respectively to the real and imaginary parts of $\rho_{01}$,
and
$s_z$, the so called population inversion, is the difference between the two 
diagonal terms $\rho_{11}-\rho_{00}$. We are mainly interested in 
$s_x$ and $s_y$ because they describe the internal coherence of the qubit.

Consider the time evolution of $\vec s$ in the presence of an  external
magnetic field $B_0$, which points in the $z$ direction.  We label  the two
eigenstates of the $\sigma_z$ spin operator with $|0\rangle$ (spin down)  and
$|1\rangle$ (spin up). These eigenstates are separated by an energy gap 
$\omega_0 = g B_0$, where $g$ is the gyromagnetic ratio. Hence the spin vector 
rotates along the $z$ axis with an angular frequency $\omega_0$ (see Fig. 1).

Let us suppose that the external noise can be modeled as a stochastic time
dependent field $B_z(t)$. Then the $xy$ component of the spin vector, i.e.
$s_x\vec{e}_x + s_y\vec{e}_y$, rotates around the $z$ axis with an 
instantaneous angular frequency $\omega_0 + \Omega (t)$, where $\Omega (t)  =
gB_z (t)$. After time $t$ the {\it average} phase factor acquired  by $s_x$ and
$s_y$ due to the presence of external noise becomes 
$\left\langle \exp \left\{ i\int_0^t \Omega (t') dt'\right\}\right\rangle$,  
where the average is to be taken over all the possible field configurations.

In order to evaluate the average over the field configurations it is convenient 
to decompose the stochastic field into a Fourier series as follows:

\begin{equation}
  B_z(t) = \sum_k b_{\bf k}^* e^{i\omega_k t} +  b_{\bf k} e^{-i\omega_k t}
\end{equation} where $b_{\bf k},b_{\bf k}^*$ are Gaussian time independent
random variables.  This allows us to write
\begin{equation}
  \left\langle \exp \left\{ i\int_0^t \Omega (t') dt'\right\} \right\rangle 
  = \prod_{\bf k} \left\langle \exp \left\{-ig \int_0^\infty b_{\bf k} 
  e^{i\omega_k t'} dt' + c.c.\right\} \right\rangle \label{<phi>}
\end{equation}

which shows how the overall phase factor can be expressed as a product  of 
independent contributions from each of the Fourier component of the 
fluctuating  field. These contributions add incoherently and destroy the  spin 
coherences. 

The process of decoherence described above is similar to what is knonw as $T_2$
decay in nuclear magnetic resonance (Allen \& Eberly 1975). This is to be 
distinguished from $T_1$ decay, which implies a damping of the spin energy 
into the field. Our model does not describe $T_1$ processes.

To illustrate the destruction of coherences in the semiclassical picture  we
simulate the energy fluctuations by explicitly introducing the random  energy
fluctuations $B_z(t)$ while calculating the time evolution of $\vec{s}$.  We do
this numerically in discrete time steps $\Delta t$. For each time step we
select a  random number $\kappa$ between 0 and 1. If $\kappa<0.1$ then 
$B_z(t+\Delta t) = B_z(t) + B'$, where $B'$ is some constant field. 
Correspondingly, if $\kappa>0.9$ then $B_z(t+\Delta t) = B_z(t) - B'$.
Otherwise $\vec{B}$ remains unchanged. We scale our variables dimensionless, 
set $B' = 0.1B_o$ and set $g=1$. In figure 2 we  show examples of such
simulations. In a more realistic model the size of  fluctuations should be
randomized as well, but this simple model with fixed-size fluctuations is good
enough for our illustration purposes. 

Each stochastic calculation of time evolution will produce a practically unique 
history for $\vec{s}$. These histories form an ensemble, and if we average over 
its all possible members, we obtain a statistical average which then corresponds
to the one shown in (\ref{<phi>}). In figure 2(a) and 2(b) we show two examples 
of single ensemble members, and in figure 2(c) we show an ensemble average of 
100 such random ensemble members. We see that as the fluctuations average out,
the changes introduced to the coherences $s_x$ and $s_y$ remain. And as seen 
in figure 2(d), with 500 members in the ensemble, they eventually  disappear
altogether as $t$ becomes large, i.e., $t\gg \omega_0^{-1}$.

Our semiclassical simulations for a single quantum bit demonstrate how the 
full statistical result is obtained by adding the individual members of the
ensemble. It should be noted that in the time evolution of  a single member of
the ensemble the amplitude for the oscillations in $s_x$ and $s_y$ remains
constant. Hence there is no real decay of the coherences  present in a single
ensemble member, but merely a shift in phase.

In the semiclassical picture it is possible to regard a single member of the
ensemble as a fully physical entity. It could describe e.g. collisionally
induced energy fluctuations in qubits formed by ions in a trap; these arise
when the atoms of the background gas collide with the ions and perturb the
ionic energy level structure. However, when we consider the effect of the
thermal and vacuum fluctuations, such a semiclassical view on the interaction
of the qubit with the reservoir  is not adequate, and one has to apply a fully
quantum approach, which we shall present next.

\subsection{Fully Quantised Model}

The above description of our decoherence mechanism can be rigorously  
formulated in quantum terms.  The quantised environment is modeled as a
continuum of field modes.  The state of the combined system i.e. qubit +
environment is described by a  density operator $\varrho (t)$ which at time
$t=0$ is assumed to be in the form

\beq
\varrho(0) = \rho (0) \otimes \prod_{\bk} R_{\bk T},
\eeq

where $R_{\bk T}$ is the thermal density matrix of the $\bk $ mode of the field
and we have taken advantage of the fact that the density operator of the field
in thermal equilibrium factorizes into the tensor product of the density 
operator of each of its modes.

We assume that the dynamics of a single qubit + environment is described   by
the following Hamiltonian 

\beq H = \half\sz \wo + \sumk \Bks \Bk \wk + \sumk  \sz (\gk\Bks + \gk^*\Bk )
\label{hamiltonian}
\eeq

where the first and the second term on the r.h.s. describe respectively    the
free evolution of the qubit and the environment, and the third term  describes
the  interaction between  the two ($\Bk , \Bks $ are now creation and
annihilation  field operators, here and in the following we put $\hbar = 1$).  
Hamiltonian (\ref{hamiltonian}) is equivalent to the one introduced in  
connection with the tunnelling problem (Leggett {\em et al.} 1987, Gardiner
1991) and used  to model decoherence in quantum computers (Unruh 1995).

Since $[\sz,H]=0$ the populations of the qubit density matrix, 
$\rho (t) = \mbox{Tr}_R  \varrho(t)$ are not affected by the environment and  
the coupling with the environment in our model simply erodes quantum
coherence.    This means that also in our fully quantised model there is  {\em
no} exchange of energy between qubit and environment and consequently  no $T_1$
type of decay takes place. This is not a limitation of the model  since, as
will became clear in the following, it describes adequately the  decoherence
mechanism which usually takes place on a shorter timescale than  the energy
dissipation.  Furthermore this model is {\em exactly} soluble and allows  a
clear analysis of the mechanism of entanglement between qubit and environment
which is believed to be at the core of most decoherence processes.

We will postpone a detailed discussion on the form of the  coupling $\gk$  
between qubit and field modes which  will depend on the specific 
characteristics of the physical system. 

In order to study the time evolution of our system it will be convenient to  
move to the interaction picture, where the time evolution operator takes  the
form

\beqa U(t) & = & \exp \left\{ -i\int_o^t  
\sumk  \sz \left( \gk\Bks \ewtp + \gk^*\Bk \emwtp \right) dt'\right\}\\ & = &
\exp \left\{ \sz \half \sumk 
\left( \Bks \xi_{\bk}(t) - \Bk \xi_{\bk}^{\ast}(t) \right) \right\}\label{U(t)}
\nonumber
\eeqa

with

\beq
\xi_{\bk}(t) =  2\gk{1 - \ewt\over \wk} .
\eeq

Here $U(t)$ can be described as a conditional displacement operator for the
field,  the sign of the displacement being dependent on the logical value of
the qubit. In particular for any pure state 
$\kett{\Psi}$ of the field

\begin{eqnarray} U(t) \kett{0}\otimes\kett{\Psi} &=& \kett{0}\otimes 
\prod_{\bk} D(-\half
\xi_{\bk}(t))\kett{\Psi}\nonumber\\ U(t) \kett{1}\otimes\kett{\Psi} &=&
\kett{1}\otimes 
\prod_{\bk}D(+\half \xi_{\bk}(t))\kett{\Psi}
\label{diss}
\end{eqnarray}

where the displacement operator $D(\xi_{\bk})$ is defined as 
\begin{equation}
 D(\xi_{\bk}) = \exp \left\{ \Bks \xi_{\bk} - \Bk \xi_{\bk}^{\ast}\right\}.
\end{equation}

The above discussion makes it clear that {\em U(t)} induces entanglement between
qubit states and field states. For example, if at time $t=0$ the state of the 
system composed by the qubit and the $\bk$ mode of the field is the   tensor
product of a general qubit state times the vacuum state for the field mode,
then the interaction Hamiltonian will, at time $t$, generate an  entangled state

\beq (c_0 |0\ket + c_1 |1\ket) \otimes |0_\bk \ket \stackrel{U(t)}{\longmapsto}
c_0 |0\ket |-\half\xi_{\bk} (t)\ket  + c_1 |1\ket |+ \half\xi_{\bk} (t)\ket ) 
\eeq

where $ |{1\over 2}\xi_{\bk} (t)\ket $ is a coherent state of amplitude
${1\over 2}\xi_{\bk }(t) $. Such correlations between qubit and environment 
cannot be accounted for in the semiclassical approach outlined  at  the
beginning of this section. It is precisely this entanglement which,  in the
fully quantised description we have adopted, is responsible for the 
decoherence process. Indeed the off-diagonal elements of the reduced density 
matrix of the qubit decay due to the fact that the overlap between the
different field states with which the qubit becomes entangled diminishes in
time.   We can formulate in rigorous terms the analysis we have outlined above
by  taking into account all the field modes in thermal equilibrium. The matrix 
elements of the reduced density operator of the qubit are  defined as

\begin{equation}
\rho_{ij} (t) = \braa{i}\mbox{Tr}_R U(t)\varrho (0) U^{-1}(t)\kett{j}.
\label{rhoij}
\end{equation}

Using Eq.(\ref{diss}) we check that, as anticipated, 
$\rho_{00}(t)=\rho_{00}(0)$, $\rho_{11}(t)=\rho_{11}(0)$ and for the coherence
$\rho_{10}$ we obtain

\beqa
\rho_{10} (t) & = & \prod_{\bk } Tr_{\bk }\{R_{\bk T} D(\xi_{\bk}(t))\}
\rho_{10}(0)
\nonumber\\ & = & e^{- \Gamma (t)} \rho_{10}(0).
\label{rho10}
\eeqa

Equation (\ref{rho10}) is {\em exact}, i.e. no approximation  has been made to
obtain it.

To evaluate $\exp\{ - \Gamma (t)\} $ we have to calculate the average value  of
the displacement operator for each mode in a thermal state. This is also known
as the symmetric order generating function for a harmonic  oscillator in
thermal equilibrium  (Hillery {\em et al.} 1984, Gardiner 1991). It can be
shown that 

\beq Tr_{\bk} \{ R_{\bk T} D(\xi_{\bk}) \} = \exp\left\{- {|\xi_{\bk}|^2\over 2}
\coth \left({\wk\over 2T }\right) \right\},
\label{trace}
\eeq

(we put the Boltzmann constant $k_B = 1$). Thus, in the continuum limit

\beqa
\Gamma (t) & \propto & \int d\bk |\gk |^2 \coth \left({\wk\over 2T }\right) {1
- \cos \wk t \over \wk^2}
 \nonumber\\ & \propto & 
\int d\w {d k \over d\w} G(\w ) |g(\w ) |^2 \left( 1 + 2\bra n (\w ) \ket_T
\right) {1 - \cos \w t \over \w^2} 
\label{gamma3}
\eeqa
 
where $G(\w )$ is the density of modes at frequency $\w $ (we have dropped the
superfluous index $k$),  
$\bra n (\w ) \ket_T  = \exp (-\w / 2T) {\rm cosech}(\w / 2T)$  is the  average
number of field excitations at temperature $T$ and $(d k / d\w)$  is the
dispersion relation.  In Eq.(\ref{gamma3}) we can separate the effects  due to
thermal noise  from the one due to purely quantum fluctuations. This formal
separation allows us to  identify the existence of various timescales in the  
decoherence process (see also Unruh 1995).  Let us first note that thermal
fluctuations can affect the qubit dynamics only for times longer  than the
characteristic thermal frequency $T$. For $t < T^{-1} $ only  vacuum quantum 
fluctuations contribute to the dephasing process.

Furthermore the quantity $G(\w )|g(\w )|^2$ is in general characterized by  a
cutoff frequency whose specific value depends on the particular nature of  the
physical qubit under investigation. For example if the  noise field   is a
phonon field the natural cutoff can be identified with the Debye frequency. 
More generally we can think of the cutoff as due to some characteristic length 
scale in our system below which  the qubit-environment coupling decreases
rapidly. Therefore we assume $(d k / d\w)G(\w )|g(\w )|^2
\propto \w^n  e^{-\w /\w_c}$. The exponent $n$ will depend on the number of
dimensions of the  field. In our analysis we will concentrate our attention to
the case of one-dimensional field, for which $n=1$ and of three-dimensional
field, for which $n=3$. It is evident that the so-called quantum vacuum
fluctuations will   contribute to the dephasing process only for times $t >
\w_c^{-1}$.

We can identify three time regimes of decoherence:

\begin{itemize}

\item a "quiet" regime, for $t<\w_c^{-1}$, where the fluctuations are 
ineffective in the decoherence process

\item a quantum regime, for $ \w_c^{-1} < t < T^{-1}$, where the main cause of
coherence loss are the quantum vacuum fluctuations

\item a thermal regime, for $t>T^{-1}$, where thermal fluctuations play the
major role in eroding the qubit coherence.

\end{itemize}

In order to have some semiquantitative picture of the time dependence of the 
decoherence we need to specify the frequency dependence  of the  density of 
states and of the coupling.  The case of one-dimensional field ($n=1$) has
already received some attention in literature, in this case

\beq
\Gamma (t) \propto  \int d\w e^{-\w / \w_c}
\coth \left({\w\over 2T }\right){1 - \cos (\w t) \over \w} .
\label{gamma1d}
\eeq

An analytic solution of (\ref{gamma1d}) can be obtained in one dimension in 
the  low temperature limit $(\w_c \gg T )$:

\beq
\Gamma (t)   \propto  \ln (1 + \w_c^2 t^2 ) +  2 \ln \left[{1\over \pi T
t}\sinh (\pi T t )\right].
\label{gamma1d'}
\eeq

The first term arises from the quantum vacuum fluctuations while the second  
is due to the thermal ones. Expression (\ref{gamma1d'}) reduces to
$\Gamma (t) \sim \w_c^2 t^2 $ for $t < \w_c^{-1}$, 
$\Gamma (t) \sim 2\ln \w_c t $ for $\w_c^{-1} < t < T^{-1}$,  and $\Gamma (t)
\sim Tt $ for $ t > T^{-1}$.

These three regimes can be easily identified in Fig.(\ref{regimes}),  which
shows the decoherence of a single qubit induced by a one-dimensional  field for
the particular choice $\w_c /T = 100 $.

It is also interesting to consider the three dimensional field case ($n=3$),
where the integral can be evaluated exactly:

\beqa
\Gamma (t)  & \propto & \int d\w\; \w e^{-\w/\w_c}
\coth \left( {\w\over 2T }\right) [ 1 - \cos (\w t) ] \nonumber\\ &\propto &
2T^2 \left\{ 2\zeta \left[ 2,{T\over\w_c}\right] - 
\zeta \left[ 2,{T\over\w_c}(1 + i\w_ct)\right] - 
\zeta \left[ 2,{T\over\w_c}(1 - i\w_ct)\right] \right\} \nonumber\\ & & +
\w_c^2 \left[{1\over (1 + i\w_ct)^2} +  {1\over (1 - i\w_ct)^2} -2 \right].
\label{gamma3d}
\eeqa

where $\zeta (x,y)$ is the generalized Riemann zeta function. In
Fig.(\ref{gamma1,3}) is shown the decay of a single qubit as a  function of
time  and as a function of the ratio $\eta = \w_c /T$ in the case of a 
one-dimensional (a) and three-dimensional (b) fields. In (a) the coherence
always decays to zero. For $\eta \sim 1$ this decay is dominated by the thermal
fluctuations of the reservoir  and it is therefore exponential. However, for
large $\eta$ there is an intermediate region where the  vacuum fluctuations
dominate and the decay is roughly linear before the  thermal regime takes over.
In all cases there is a short $t$ region  where almost no decay is present. As
$\eta$ decreases, the decay onset moves towards larger $t$. This is because the
cut-off frequency $\w_c$  determines the extent of the "quiet" regime. In (b)
the system shows  a very different behaviour: the decoherence saturates to a
value determined by $\eta$. This difference in behaviour is reminiscent of the
dimensional differences observed in various models of phase transitions (Ma
1976)  and is due to the suppressed influence of low frequency fluctuations in
three dimensions. It should however be pointed out that at longer timescales,
when $T_1$ processes involving exchange of energy between qubit and environment
take place, this residual coherence will disappear.

\section{Decoherence of Quantum Registers}

In the present literature on decoherence processes in quantum computation  it is
usually assumed that in a register of length $L$ each qubit interacts 
individually with a different reservoir. In this case all the analysis of the
relevance of dissipation from the complexity viewpoint is done assuming that
$\Gamma_{L}(t) \sim L \Gamma(t)$  (see next section). In this section we
discuss in which circumstances this  assumption is justified and we analyse the
consequences of collective  interaction on  the complexity analysis, when such
effects need to be  taken into account. 

We will start by considering a system of two qubits at positions 
$\br_a , \br_b$ respectively. Their dynamics is described by the  following
Hamiltonian

\beqa H & = & \half\sza \woa +\half  \szb \wob +\sumk \Bks \Bk \wk \nonumber\\
 & + & \sumk \left( \sza  (\gka\Bks + \gkas\Bk )
 +  \szb  (\gkb\Bks + \gkbs\Bk )\right).
\label{hamiltonian2}
\eeqa 

Hamiltonian (\ref{hamiltonian2}) is a straightforward generalization of 
(\ref{hamiltonian}) where $\gka , \gkb $ are position dependent couplings. In
(\ref{hamiltonian2})  we are not considering any direct interaction between the
two qubits which  would be necessary for the conditional dynamics (Barenco {\em
et al} 1995). Here we concentrate our attention to the simplest case of a 
system of two  qubits coupled only to the environment.

In the interaction picture the time evolution operator takes the form

\beq U(t) =  \exp \left\{ \half \sumk 
\Bks \left[ \sza \xi_{\bk}^a(t) + \szb \xi_{\bk}^b(t) \right] -
\Bk \left[ \sza \xi_{\bk}^{a*}(t) + \szb \xi_{\bk}^{b*}(t) \right] \right\}.
\eeq

Here again the unitary evolution produces entanglement between register states 
and environment states. It should be noted however that $U(t)$ acts as a  
displacement operator on the field with a displacement amplitude depending on 
the logic value of {\em both}  qubits of the register. The following simple 
example will help to clarify this point. Let us consider the following two 
initial states

\beqa |\Phi^{(-)} \ket & = & ( c_{10} |1_a , 0_b \ket + c_{01} |0_a , 1_b\ket )
\otimes | 0_{\bk} \ket,\\ |\Phi^{(+)} \ket & = & ( c_{00} |0_a , 0_b \ket +
c_{11} |1_a , 1_b\ket )
\otimes |0_{\bk}\ket,
\eeqa

where we consider for simplicity only the vacuum state of the 
$\bk $ mode and where the $c_{ij}$ are arbitrary complex amplitudes.  Due to
the qubit-field interaction the system will evolve into the states 

\beqa |\Phi^{(-)} (t) \ket & = &  c_{10} |1_a , 0_b \ket |+\half (\xi_{\bk}^a -
\xi_{\bk}^b ) \ket + c_{01} |0_a , 1_b\ket |\half(\xi_{\bk}^a - \xi_{\bk}^b )
\ket ,\\ |\Phi^{(+)} (t)\ket & = &  c_{00} |0_a , 0_b \ket |-\half (\xi_{\bk}^a
+ \xi_{\bk}^b )\ket + c_{11} |1_a , 1_b\ket |\half (\xi_{\bk}^a + \xi_{\bk}^b
)\ket .
\eeqa

Since the amount of decoherence is measured by the overlap between the two 
different field states with which the qubit states become entangled,  the
states $|\Phi^{(+)}\ket, |\Phi^{(-)}\ket $ will be characterized by  different
decoherence times. In particular, in the limit 
$\br_a \approx \br_b$  we have $\xi_{\bk}^a \sim \xi_{\bk}^b$ and  therefore 
in the "subdecoherent" state $|\Phi^{(-)}\ket$ the qubits are not entangled 
with the field, while in the "superdecoherent" state $|\Phi^{(+)}\ket $ the
qubits are collectively entangled with a coherent  state with an amplitude 
which is twice that of the single qubit case.

We can formulate in more rigorous terms the analysis we have just sketched 
above by taking into account all the modes of the field in thermal equilibrium. 
We will again concentrate our attention on the reduced density matrix of the 
two qubit system whose matrix elements can be conveniently expressed as

\beq
\rho_{i_aj_a,i_bj_b}(t) = \bra i_a , i_b | Tr_R\{\varrho (t)\}| j_a , j_b \ket .
\eeq

It is straightforward to verify that the analogous of Eq. (\ref{rho10}) is

\beq
\rho_{i_aj_a,i_bj_b}(t) = \rho_{i_aj_a,i_bj_b}(0)
\prod_{\bk} Tr_{\bk} \left\{ R_{\bk T}  D[(i_a - j_a)\xi_{\bk}^a]D[(i_b -
j_b)\xi_{\bk}^b] \right\} .\nonumber
\label{rholmpq}
\eeq

Matrix elements of the form $\rho_{i_ai_a,i_bj_b}$ show no collective decay, 
in fact

\beq
\rho_{i_a i_a,i_bj_b}(t) = \rho_{i_ai_a,i_bj_b}(0)
\prod_{\bk} Tr_{\bk} \left\{ R_{\bk T}  D[(i_b - j_b)\xi_{\bk}^b] \right\}
\nonumber
\eeq

which for $i_b = j_b$ reduces to 
$\rho_{i_a i_a,i_bi_b}(t) = \rho_{i_ai_a,i_bi_b}(0)$ while for $i_b \neq j_b$
we have 
$\rho_{i_a i_a,i_bj_b}(t) = \rho_{i_ai_a,i_bj_b}(0)e^{-\Gamma (t)}$. Analogous
results hold of course for matrix elements of the form
$ \rho_{i_aj_a,i_bi_b}$.

Collective decay is instead apparent in the matrix elements of the form
$\rho_{i_a j_a,i_bj_b}(t) $ with $i_a \neq j_a, i_b \neq j_b$ :

\beqa
\rho_{10,10}(t) & = & \rho_{10,10}(0)
\prod_{\bk} Tr_{\bk} \left\{ R_{\bk T} D(\xi_{\bk}^a + \xi_{\bk}^b) \right\} 
\nonumber\\ & = & e^{-\Gamma^+(t)}\rho_{10,10}(0),\\
\rho_{10,01}(t) & = & \rho_{10,01}(0) 
\prod_{\bk} Tr_{\bk} \left\{ R_{\bk T} D(\xi_{\bk}^a - \xi_{\bk}^b) \right\} 
\nonumber\\ & = & e^{-\Gamma^-(t)}\rho_{10,01}(0).
\eeqa

The  expression for $\Gamma^{\pm}$ can obtained from (\ref{trace}) with the 
obvious substitution $\xi_{\bk}\rightarrow \xi_{\bk}^a \pm \xi_{\bk}^b$.  If we
suppose $ \gk^a = \gk e^{i\bk\cdot\br_a}$ and  
$ \gk^b = \gk e^{i\bk\cdot\br_b}$ , then, in the continuum limit, the
collective decay rates are

\beq
\Gamma^{\pm} (\bR , t) \propto  \int d\bk |g_{\bk}|^2  
\coth \left({\wk\over 2T }\right) {1 - \cos \wk t \over \wk^2} [ 1 \pm \cos(\bk
\cdot \bR)]
\label{gammapm}
\eeq

where $\bR = \br_a - \br_b$.

The origin of the position dependence of the collective decay constants 
$\Gamma^{\pm}(\bR, t)$ can be easily explained both in semiclassical and in 
quantum terms. We have already mentioned that in a system of two qubits the 
entanglement with the field will in general depend on the qubit separation.  In
a semiclassical picture if the distance between the two qubits is smaller  than
the correlation length of the bath  they ``feel" the same  fluctuation.  More
generally the random phase-shifts due to the bath fluctuations of wavelength
shorter than the inverse of the qubit separation are the same for  both qubits. 
This produces a reduction - or an enhancement - of the decay rates depending 
on whether the phase shifts on the two qubits add or subtract.

In order to avoid possible confusions it must be stressed that the  phenomenon
of superdecoherence vs. subdecoherence that we have just described is {\em not}
the same as the process of superradiance vs. subradiance more commonly
encountered in litterature (see e.g Allen \& Eberly 1975). While
superdecoherence is due  to {\em collective entanglement} between qubits and
environment with no  exchange of  energy, superradiance is a process of {\em
collective radiation}  by a group of  closely spaced atoms. Indeed the
conditions for the two processes to occur are quite different: superdecoherence
requires the qubit to be  in a region of space smaller than the coherence
length of the environment, which is a much more stringent condition that the
one needed for  superradiance to  take place, i.e. that the atoms are at a
distance smaller  than the  wavelength of the resonant field modes.
Furthermore, it must be noted  that some superradiant states are subdecoherent.
In general the set of  states which are robust against the dephasing action of
the environment will depend on the specific form of qubit-environment coupling.

In the one-dimensional and three-dimensional field cases, the collective  decay
constants are, respectively, given by  

\beq
\Gamma^{\pm}_{1D}(\bR ,t) \propto  2\int_0^\infty d\w e^{-\w / \w_c} \coth
\left({\w\over 2T }\right) {1 - \cos \w t \over \w} [ 1 \pm \cos(\w t_s)]
\label{gpm1d}
\eeq
\beq
\Gamma^{\pm}_{3D}(\bR , t) \propto 2\int_0^\infty d\w \w e^{-\w / \w_c}\coth
\left({\w\over 2T }\right) (1 - \cos \w t ) \left( 1 \pm {\sin(\w t_s)\over\w
t_s }\right)
\label{gpm3d}
\eeq

where we have introduced the transit time 
$ t_s $ ($\w t_s = \bk \cdot \bR$).

The analysis carried out for a two-qubit register can be easily extended  to
the  case of a register of $L$ qubits, whose density matrix elements can be
written as
 
\beqa
\rho_{\{i_n,j_n\}}(t) & = &
\bra i_{l-1},i_{L-2} ... , i_{0} | Tr_R\{\varrho (t)\}| j_{L-1},j_{L-2},...,
j_{0}\ket \\
 & = & 
\rho_{\{i_n,j_n\}}(0)\nonumber\\ & \times &\prod_{\bk} Tr_{\bk} \left\{ R_{\bk
T}  D[(i_{L-1} - j_{L-1})\xi_{\bk}^{(L-1)}]D[(i_{L-2} -
j_{L-2})\xi_{\bk}^{(L-2)}] \right.\nonumber\\ & & \left. ....D[(i_{0} -
j_{0})\xi_{\bk}^{(0)} \right\} .\nonumber
\eeqa

From the viewpoint of the following complexity analysis it is interesting to 
consider the limiting  case in which all the qubits of the register are  in the
same position. In this case all the $\xi_{\bk}^{(i)}$ are equal and  the matrix
elements with the fastest decay are
$\rho_{\{1_n,0_n\}}$ and $\rho_{\{0_n,1_n\}}$ for which we have

\beq
\rho_{\{1_n,0_n\}} = \rho_{\{1_n,0_n\}}(0) e^{-L^2\Gamma (t)}.
\eeq 

In general,

\beqa
\rho_{\{i_n,j_n\}}(t) & = & 
\rho_{\{i_n,j_n\}}(0)\prod_{\bk} Tr_{\bk} \left\{ R_{\bk T} 
D\left[\sum_{n=0}^{L-1}(i_n - j_n)\xi_{\bk}\right] \right\}\nonumber\\ & = &
\rho_{\{i_n,j_n\}}(0) \exp \left\{ - \left|\sum_{n=0}^{L-1} (i_n -
j_n)\right|^2 \Gamma (t) \right\} 
\label{same}
\eeqa

which should be compared with the expression for the decoherence in the case of
independent reservoirs

\beqa
\rho_{\{i_n,j_n\}}(t)& = & \rho_{\{i_n,j_n\}}(0) \exp \left\{ -\sum_{n=0}^{L-1}
\left| (i_n - j_n)\right| \Gamma (t) \right\} 
\label{independent}
\eeqa

where $\sum_{n=0}^{L-1}|(i_n - j_n)| $  is the Hamming distance between the two
qubit states.  In the independent reservoir case therefore in the worst case
the density  matrix elements decay as $\exp \{-L\Gamma (t)\}$.

\section{Discussion and Conclusions}

Our analysis shows that decoherence destroys quantum interference and 
entanglement in quantum computers, thus decreasing the probability of 
successful computation exponentially with the input size $L$.   Reservoirs with
large coherence lengths  introduce only asymmetry in the  decoherence rates of
various off-diagonal elements of the density matrix with the same Hamming
distance between its indices (compare Eq. (\ref{same}) with Eq.
(\ref{independent})).  In this case while some coherences are not affected by
the decoherence some others are destroyed more rapidly. Assuming that all
coherences are  {\em equally important} and that for the complexity analysis we
choose the  worst case (i.e. the off-diagonal element which decays at the
fastest rate)  then we are forced to conclude that reservoirs with a large
coherence length   can be more of a hindrance than a help.  In the independent
reservoirs limit the fastest decoherence is proportional to $\exp(-L\Gamma
(t))$ whereas for the reservoirs with a large  coherence length there are
off-diagonal elements which decay as 
$\exp(- poly(L)\Gamma (t))$. However, if one can support quantum computation 
with some selected coherences then a large coherence length can be utilised.

Let us suppose that we can manufacture in our laboratory a quantum register of
2L qubits composed of pairs of qubits close enough to each other so that each
pair is effectively interacting with the same reservoir. Different pairs can
interact with different reservoirs although the results we are going to 
illustrate are not modified if all the qubits interact with the same reservoir.
The idea is that if we use a pair of qubits to encode a bit we might 
effectively decouple the register from its environment. To give a simplified
example of how this might work let us consider a register of 4 qubits in which
the two pairs of qubits interact with two different reservoirs. The interaction
will evolve the system as 

\beq |i_a , j_a \ket |i_b , j_b \ket |0^a_{\bk} \ket |0^b_{\bk}\ket
\longrightarrow |i_a , j_a \ket |i_b , j_b \ket  |(-1^{i_a}
-1^{j_a})\xi^a_{\bk} \ket |(-1^{i_b}-1^{j_b})\xi^b_{\bk}\ket
\eeq

where for simplicity we have considered only the vacuum state of one mode for
each reservoir, labelled {\it a}, {\it b}, respectively. It is obvious that
states for  which $i_a \neq j_a$ and 
$i_b \neq j_b£$  do not get entangled with the reservoir and therefore they
mantain their coherence. We might therefore use the following encoding 

\beqa |\tilde 0\ket = |0,1\ket , \nonumber\\ |\tilde 1\ket = |1,0\ket .
\eeqa

This "noiseless" encoding does not pose complexity problems since to encode L
qubits we need simply 2L qubits. Of course, the experimental realization of a
register with the above characteristics encounters practical problems; it
might, for instance, be difficult to  excite and manipulate only the stable
coherences.  The main point we want to make, however, is that appropriate
encoding  could be a way to overcome some of the problems posed by couplings
with  external noise.

Is the computational complexity analysis dependent on the character of
decoherence i.e. on the type of coupling with the environment? We believe  that
as long as the reservoir is in a thermal state, any model of decoherence  will
lead to the same conclusions. In the independent reservoirs limit the fastest
decoherence is proportional to
$\exp(-L\Gamma(t))$ regardles the functional dependence $\Gamma(t)$. For
reservoirs with a large coherent length the fastest decoherence will be
proportional to
$\exp(-poly(L)\Gamma'(t))$ leading to the same unwelcome exponential increase
of the error rate. However, let us emphasise again that this should not be
taken  as a fundamental argument against possibilities of experimental
implementation of quantum factorisation. Quantum error correction, modification
of the existing quantum algorithms, customizing the environmental noise, and
probably many other techniques can be employed to improve the efficiency of
quantum algorithms.

Even if quantum algorithms for factorisation are not efficient from the
complexity theory point of view, they may still permit factoring of numbers
which are  much bigger than those that can be factored using classical
algorithms.  Technological progress in isolating quantum computers from the
environment and suppressing decoherence will increase the size of the biggest
number  that can be factored by such computers. For if the elementary 
computational step takes time $\tau$ and $t$ is a decoherence time of a  single
qubit then the requirement for the coherent computation to be completed  within
the decoherence time of the computer can be written as

\begin{equation}
\tau L^2 < t/L.
\end{equation} 
  
The l.h.s. represents the time needed to factor number of size $L$ with the 
Shor algorithm and the r.h.s. represents the decoherence time of $L$ qubits 
(in the independent reservoirs approximation).  From the equation above follows
that the size
$L$ is bounded by 

\begin{equation} L<\left(\frac{t}{\tau}\right)^{1/3}.
\end{equation}

Thus the ratio $t/\tau$, which depends on the technology employed,  determines
the limits of the algorithm and it is unrealistic to assume that  this ratio
can be made infinite. For a review of possible values of $t$ and $\tau$ see
DiVincenzo 1995. 

Clearly the decoherence problem calls for a satisfactory solution if we want to
make quantum factoring realistic and eventually practical. The authors believe
that an efficient quantum error correction is needed in order to achieve this
goal.

\section{Acknowledgments} The authors wish to thank Adriano Barenco, Keith
Burnett, David Deutsch and David DiVincenzo for various stimulating discussions
on the subject of this paper. Part of this work has been carried out during the
ISI Workshop on Quantum Computation (Torino 1994) sponsored  by  ELSAG-BAILEY
Genova.  Financial support from The Royal Society, London, the U.K. 
Engineering and Physical Sciences Research Council,  and the European Union HCM 
Programme is acknowledged.

\begin{figure}[ht]
\caption[f1]{The Bloch vector $\vec{s}$ and its projections in the Cartesian
coordinate system. The field vector $B_0\vec{e}_z$ lies in the {\it
z}-direction. When $B_0$ is constant the Bloch vector simply rotates around the
field vector and its tip follows a closed, circular trajectory, which is
parallel to the {\it xy}-plane. If the qubit is in an equal superposition of
states 0 and 1, then $s_z = 0$ and the Bloch vector itself lies in the {\it
xy}-plane.}
\end{figure}

\begin{figure}[ht]
\caption[f2]{Results from a semiclassical stochastic simulation.  In (a) and
(b)  the time evolution of $s_x$ and $s_y$ (solid lines) generated by the 
fluctuating field $B_z$ is shown for two  different ensemble members. The
dotted lines represent the unperturbed evolution. Time $t$ is in units of
$\omega_0^{-1}$. We have chosen $c_0=c_1=2^{-1/2}$  (i.e. $s_z=0$). In (c) and
(d) we show averages over 100 and 500 ensemble  members, respectively. $B_z$
averages rapidly to zero as the ensemble size increases.}
\end{figure}

\begin{figure}[ht]
\caption[f3]{The interaction between qubit and reservoir causes a  conditional
displacement of the field mode. The sign of the displacement depends on the
qubit state.  In the displaced modes the field state oscillates with amplitude
$\pm\half \xi_{\bk}$. This conditional displacement causes a qubit-field
entanglement.}
\end{figure}

\begin{figure}[ht]
\caption[f3b]{Decoherence of a single qubit for a one-dimensional field density
of states for $\w_c / T = 100$ . Time is in units of $T^{-1}$ and the
proportionality factor is equal to 0.1. The three decay regimes can be easily
identified.}
\label{regimes}
\end{figure}

\begin{figure}[ht]
\caption[f4]{Decoherence of a single qubit as a function of time and  of the
ratio  $\eta = \omega_c/T$. Time is in units of $T^{-1}$ and the 
proportionality factor has been set equal to 0.1. Here (a) shows the result of
a numerical integration of Eq.~(\ref{gamma1d}) for the one-dimensional density
of states while (b) shows of the exact solution for the three-dimensional
density of states (Eq.\ref{gamma3d}) }
\label{gamma1,3}
\end{figure}

\begin{figure}[ht]
\caption[f5]{The decay of two qubit coherence in the case of the shared 
reservoir with one-dimensional density of states (Eq.~\ref{gpm1d}). We have 
set $\eta=1$ and the proportionality factor is chosen such that at the limit of
large $t_s$ we get the results of Fig. 5(a). In (a) we see  how the decay is
cancelled out , and in (b) we see how it is amplified , when
$t_s$ is small. The onset of decay does not change with $t_s$.}
\end{figure}

\begin{figure}[ht]
\caption[f6]{The decay of two qubit coherence in the case of the shared 
reservoir with three-dimensional density of states (Eq.~\ref{gpm3d}). We have 
set $\eta=1$ and the proportionality factor is chosen such that at the limit of
large $t_s$ we get the results of Fig. 5(b). In (a) we see  how the decay is
cancelled out, and in (b) we see how it is amplified, when
$t_s$ is small. The saturation of the decay is present in both cases, though.}
\end{figure}

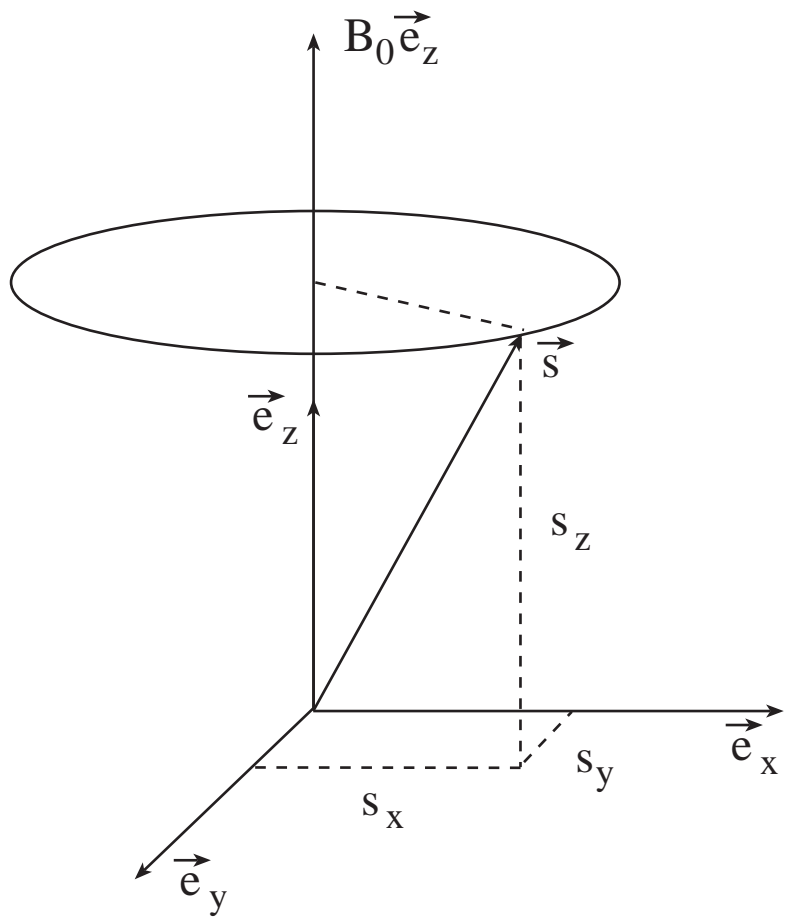
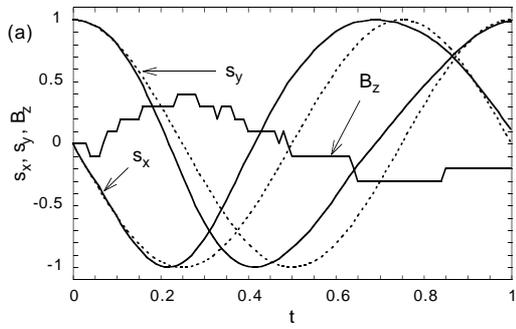
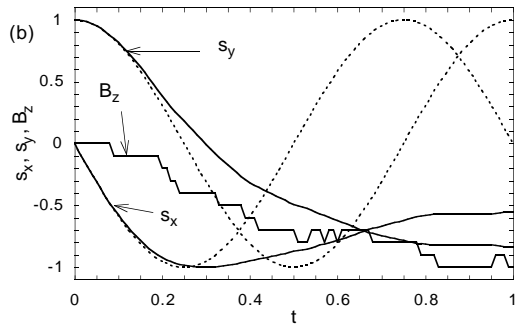
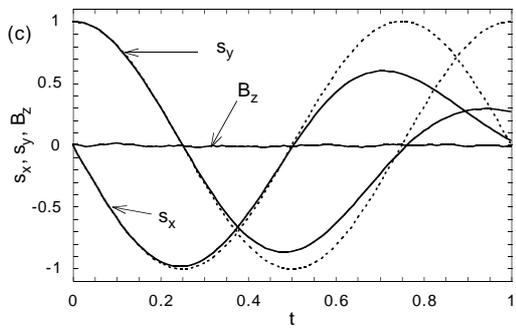
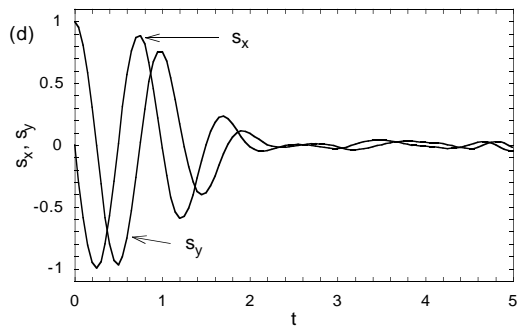
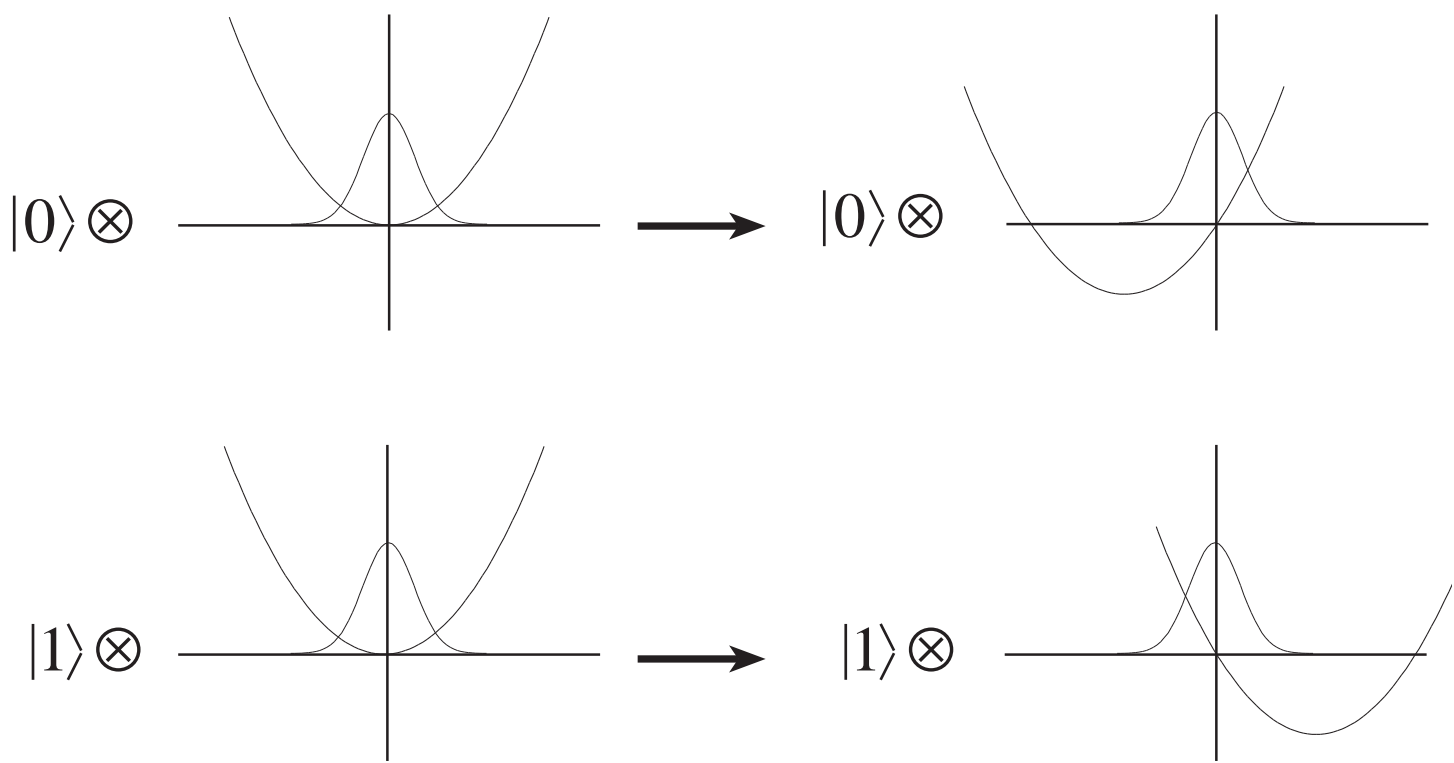
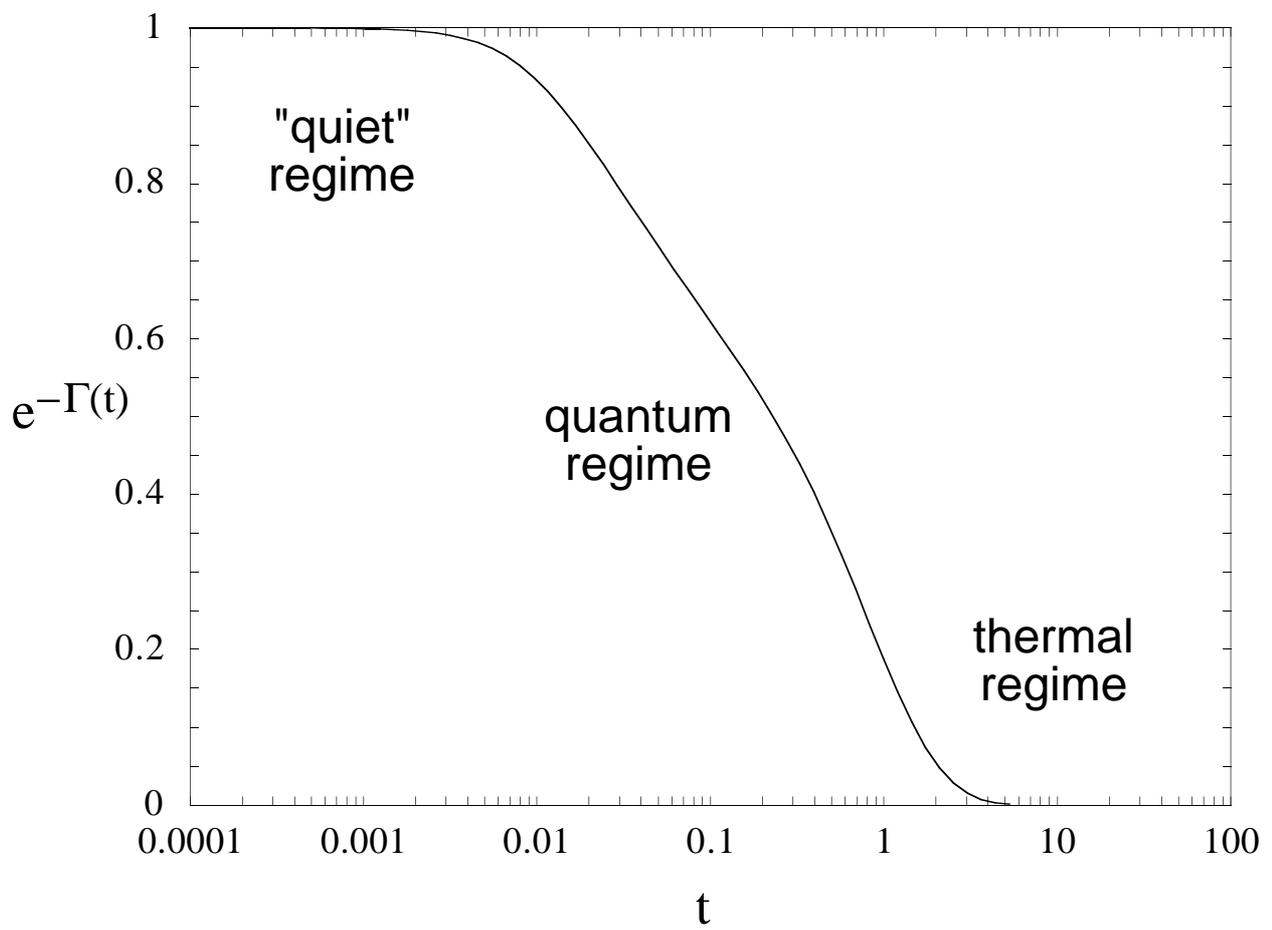
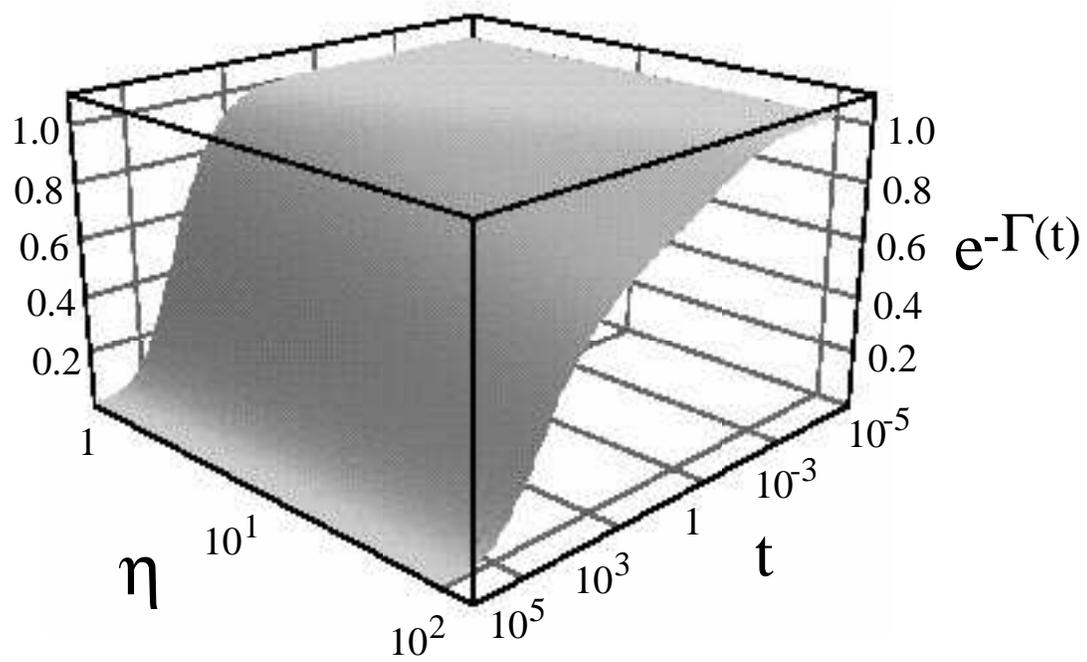
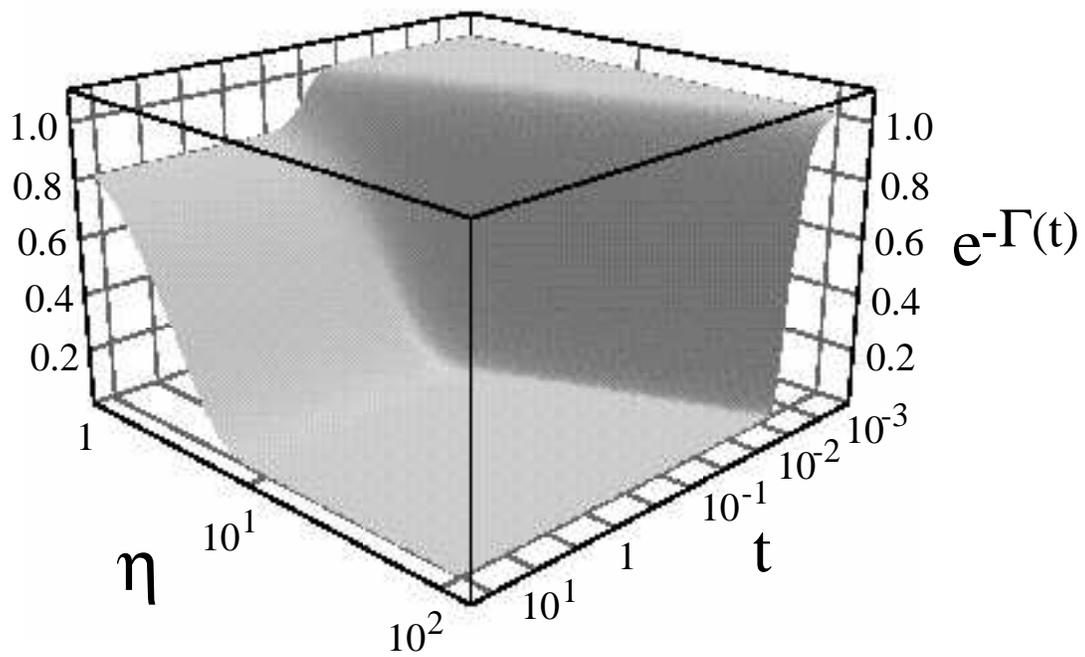
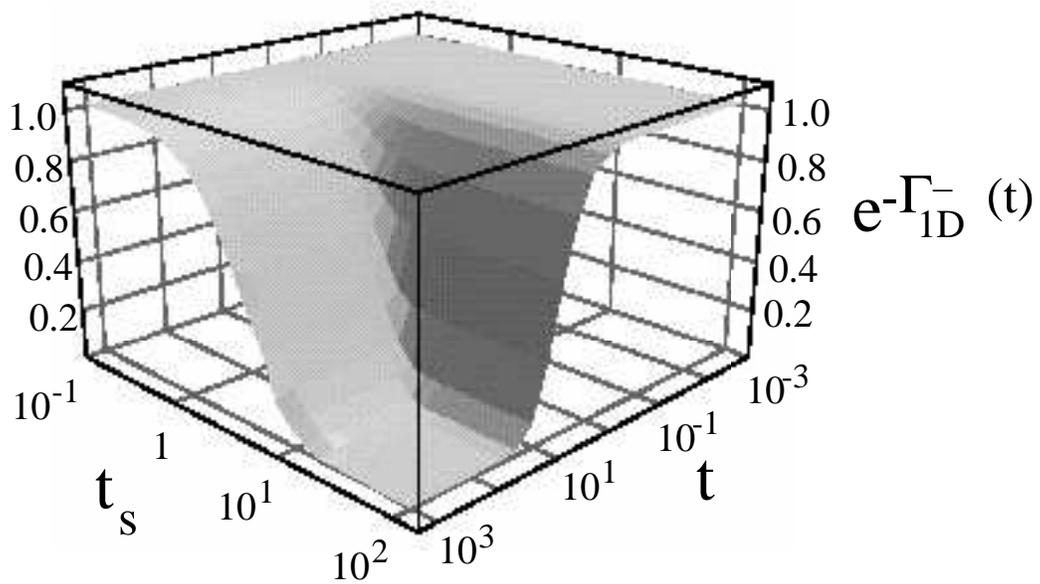
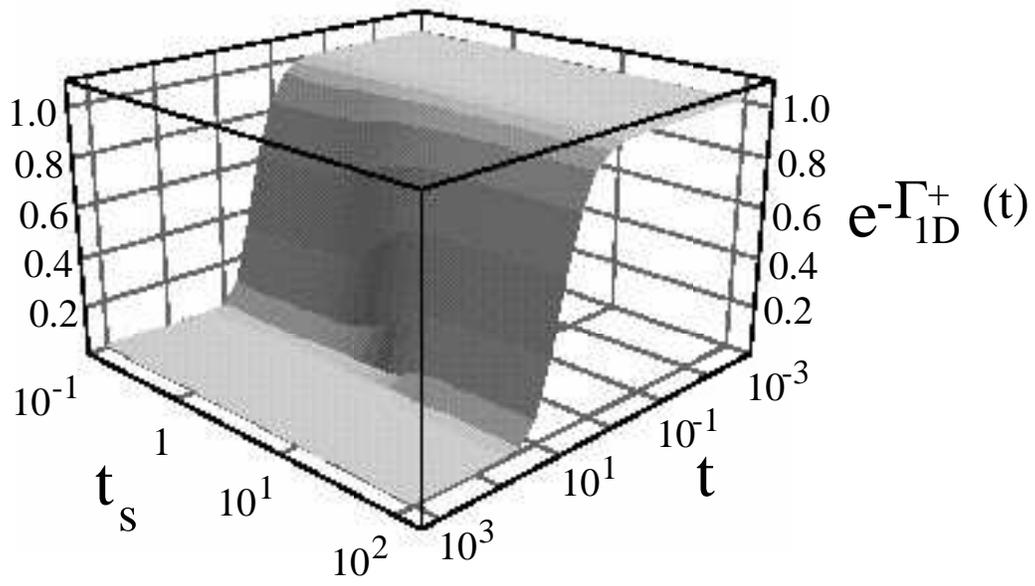
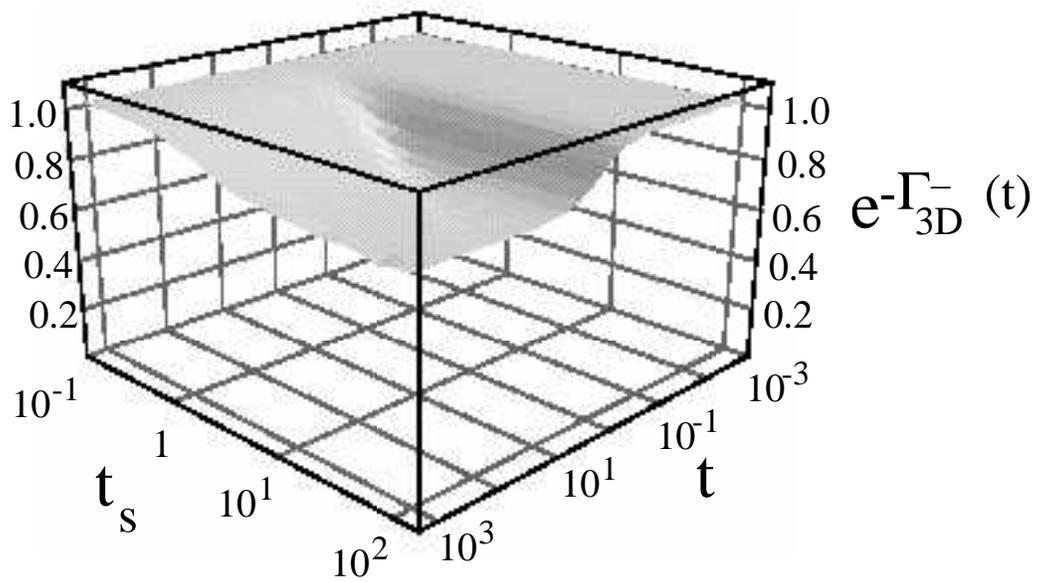
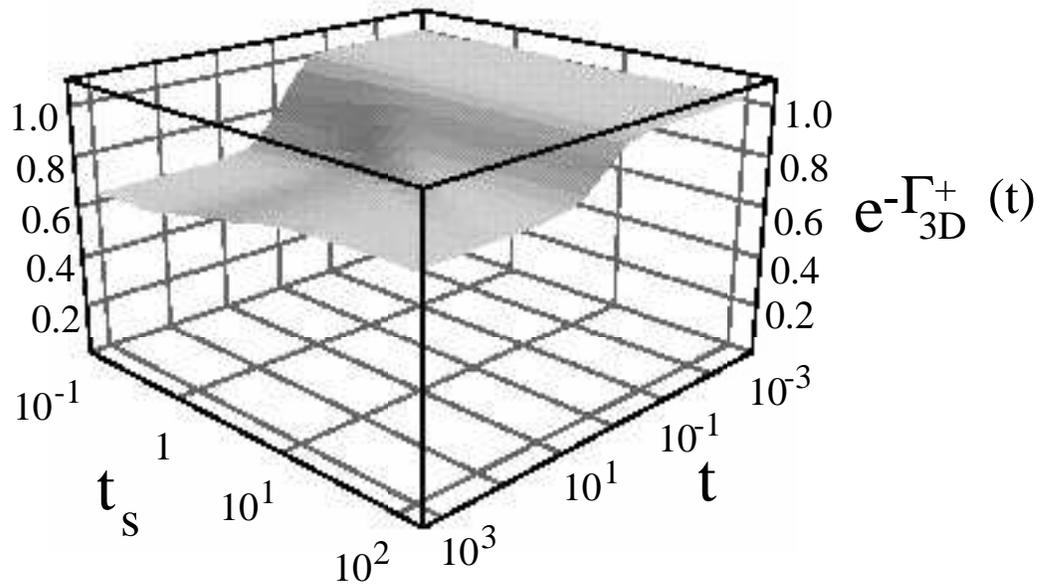
\end{document}